\documentclass[12pt]{article}
\usepackage{amsbsy}

\newcommand{\sect}[1]{\setcounter{equation}{0}\section{#1}}

\newcommand{\eq}{\begin{equation}}
\newcommand{\eqa}{\begin{eqnarray}}
\newcommand{\en}{\end{equation}}
\newcommand{\ena}{\end{eqnarray}}
\newcommand{\enn}{\nonumber \end{equation}}

\def\sk{\vskip .4cm}
\def\noi{\noindent}
\def\om{\omega}

\def\ga{\gamma}

\let \part\partial

\def\unquarto{{1 \over 4}}

\def\unmezzo{{1 \over 2}}
\def\epsi{\varepsilon}
\def\we{\wedge}

\def\de{\delta}

\def\part{\partial}

\def\sk{\vskip .4cm}

\def\noi{\noindent}

\def\X0{X^0}

\def\om{\omega}

\def\ga{\gamma}

\def\unquarto{{1 \over 4}}
\def\unmezzo{{1 \over 2}}
\def\epsi{\varepsilon}
\def\epsibold{{\bf \epsilon}}

\def\we{\wedge}

\def\de{\delta}

\def\Rcal{{\cal R}}

\def\square{{\,\lower0.9pt\vbox{\hrule \hbox{\vrule height 0.2 cm
\hskip 0.2 cm \vrule height 0.2 cm}\hrule}\,}}

\def\epsilonbar{{\bar \epsilon}}

\def\psibar{\bar \psi}

\def\rhobar{\bar \rho}

\def\Om{\Omega}
\def\Qbar{\bar Q}

\def\Sigmabar{\overline \Sigma}

\def\Rbold{{\bf R}}

\def\Ombold{{\bf \Om}}

\def\epsibold{\boldsymbol {\epsilon}}

\def\Gammabold{{\bf \Gamma}}

\begin{document}

\begin{titlepage}
\rightline{ARC-17-04}
\vskip 2em
\begin{center}
{\Large \bf Gauge supergravity in $D=2+2$  } \\[3em]

\vskip 0.5cm

{\bf
Leonardo Castellani}
\medskip

\vskip 0.5cm

{\sl Dipartimento di Scienze e Innovazione Tecnologica
\\Universit\`a del Piemonte Orientale, viale T. Michel 11, 15121 Alessandria, Italy\\ [.5em] INFN, Sezione di 
Torino, via P. Giuria 1, 10125 Torino, Italy\\ [.5em]
Arnold-Regge Center, via P. Giuria 1, 10125 Torino, Italy
}\\ [4em]
\end{center}

\begin{abstract}
\sk

We present an action for chiral $N=(1,0)$ supergravity in $2+2$ dimensions. 
The fields of the theory are organized into an $OSp(1|4)$ connection supermatrix, and are given by the 
usual vierbein $V^a$, spin connection $\omega^{ab}$,  and Majorana gravitino $\psi$.  In analogy with a 
construction used for $D=10+2$ gauge supergravity, 
the action is given by $\int STr ({\bf R}^2 {\bf \Gamma})$, where ${\bf R}$ is the  $OSp(1|4)$ curvature supermatrix two-form, and ${\bf \Gamma}$ a constant supermatrix
containing $\gamma_5$. It is similar, but not identical to the MacDowell-Mansouri action for $D=2+2$ supergravity.  The constant supermatrix breaks $OSp(1|4)$ gauge invariance to a
subalgebra $OSp(1|2) \oplus Sp(2)$, including a Majorana-Weyl supercharge. Thus half
of the  $OSp(1|4)$ gauge supersymmetry survives. 
The gauge fields are the selfdual part of $\omega^{ab}$ and
the Weyl projection of $\psi$ for $OSp(1|2)$, and the antiselfdual part of  $\omega^{ab}$ for $Sp(2)$. Supersymmetry transformations, being part of a {\it gauge} superalgebra, close off-shell. The 
selfduality condition on the spin connection can be consistently imposed, and the
resulting ``projected" action is $OSp(1|2)$ gauge invariant.

\end{abstract}

\vskip 4cm \noi \hrule \vskip .2cm \noi {\small
leonardo.castellani@uniupo.it}

\end{titlepage}

\newpage
\setcounter{page}{1}

\tableofcontents

\sect{Introduction}

In most approaches to supergravity, local supersymmetry appears as the ``square root" of
diffeomorphisms, and has a natural interpretation 
as coordinate transformation along Grassmann directions.  In this framework supersymmetry is part of
a superdiffeomorphism algebra in superspace. 

In Chern-Simons supergravities, on the other hand, supersymmetry ``lives" on the fiber of a gauge supergroup rather than on a (super) base space. It is part of a {\it gauge} superalgebra of 
transformations leaving the Chern-Simons action invariant, up to boundary terms.

These two conceptually different ways of interpreting supersymmetry are fused together
in the group geometric approach (a.k.a. group manifold or rheonomic framework, see for ex.
\cite{groupmanifold}). Recent advances in superintegration theory \cite{Castellani2014} have shown how this approach interpolates between superspace and component actions. 

In this paper we work in the gauge supersymmetry paradigm, that has been explored since long ago \cite{Chamseddine1977,MDM,PvN2} and has allowed the construction of Chern-Simons supergravities in odd dimensions \cite{Chamseddine1989,Troncoso1998,Zanelli2005}. Recently it has been used to construct chiral gauge supergravity in $D=10+2$ dimensions \cite{Castellani2017b}. The twelve dimensional action
is written in terms of the $OSp(1|64)$ curvature supermatrix, but is invariant only under its
 $OSp(1|32) \oplus Sp(32)$ subalgebra. Supersymmetry is part of this superalgebra: it is
 generated by a Majorana-Weyl supercharge and closes off-shell.
The constructive procedure relies on the existence of Majorana-Weyl fermions, and can in principle be applied in all even dimensions with signatures $(s,t)$ satisying  $s-t=0~ (mod~8)$.

Here we apply it for the case $s=2,t=2$ to find an action for $D=2+2$ 
chiral supergravity. This action is given by $\int STr ({\bf R} \wedge {\bf R}  {\bf \Gamma})$,
where ${\bf R}$ is the $OSp(1|4)$ curvature supermatrix two-form, and ${\bf \Gamma}$ is
a constant supermatrix involving $\gamma_5$. Due to the presence of ${\bf \Gamma}$, the
action is not invariant under the full $OSp(1|4)$ superalgebra, but only under a subalgebra
$OSp(1|2) \oplus Sp(2)$ that includes a Majorana-Weyl supercharge. Thus chiral (1,0) supersymmetry
survives. This is an important difference with the MacDowell-Mansouri action, for which
gauge supersymmetry is completely broken \footnote{it is ``restored" in
second order formalism, or by modifying the 
spin connection transformation law, see for ex. \cite{PvN2,Castellani2013}.}

Supergravity theories in $D=2+2$ dimensions have been considered by many authors in the past (for
a very partial list of references see \cite{Ketov1992,Bergshoeff1992}). They are 
candidate backgrounds for the $N=2$ superstring \cite{DAdda1987,Ooguri1990,deBoer1997}, and are related after dimensional reduction to integrable models in  $D=2$ \cite{Ward1985,Hitchin1986,Gates1992}.
The actions were obtained in most cases by supersymmetrizing the self-dual Einstein-Palatini action with (2,2) signature, supersymmetry invariance being of the ``base space" type, as in usual supergravity in $D=3+1$. The version we propose here differs because supersymmetry is 
{\sl chiral} (1,0), and  part of a {\it gauge} superalgebra, entailing automatic off-shell closure
 without need of auxiliary fields. 
 
The paper is organized as follows. Section 2 recalls the definitions of 
$OSp(1|4)$ connection and curvature, and their $5 \times 5$ supermatrix representation. Section 3 deals with the chiral $D=2+2$ action, its invariances, field transformation laws, the explicit expression of the action in terms of component fields, field equations and selfduality condition. Section 4 contains some conclusions. Gamma matrix conventions in $D=2+2$ are summarized in the Appendix.

\sect{$OSp(1|4)$ connection and curvature}

 This section and the next one closely parallel
the analogous sections for $D=10+2$ supergravity of ref. \cite{Castellani2017b}. 

\subsection{The algebra }

\noi The $OSp(1|4)$ superalgebra is given by:
\eqa
& & [M_{ab},M_{cd}] =  \eta_{bc} M_{ad} + \eta_{ad} M_{bc} - \eta_{ac} M_{bd} - \eta_{bd} M_{ac}  
\label{MMcomm}\\
& & [M_{ab},P_{c}] = \eta_{bc} P_{a} - \eta_{ac} P_{b}  \label{MPcomm}\\
& & [P_{a},P_{b}] = M_{ab} 
\label{PPcomm}\\
 & & \{ \Qbar_{\alpha}, \Qbar_{\beta} \} = -(C\gamma^a)_{\alpha\beta} P_{a} + {1 \over 2}
 (C\gamma^{ab})_{\alpha\beta} M_{ab}
  \label{QQcomm} \\
   & & [M_{ab},\Qbar_{\beta}]={1 \over 2} ~(\gamma_{ab})^{\alpha}_{~\beta} \Qbar_\alpha \label{MQcomm}\\
  & & [P_{a},\Qbar_{\beta}]={1 \over 2} ~(\gamma_{a})^{\alpha}_{~\beta} \Qbar_\alpha \label{PQcomm}
 \ena
where $M_{ab}$ and $P_a$, dual to the one-forms $\omega^{ab}$ (spin connection) and $V^a$ (vierbein), generate the $Sp(4) \approx SO(3,2)$ bosonic subalgebra, and the 
supercharge $\Qbar_{\alpha}$ is dual to the Majorana gravitino $\psi^\alpha$. Conventions
on $D=2+2$ gamma matrices and charge conjugation $C_{\alpha\beta}$ are given in the Appendix.

\subsection{The 5 $\times$ 5 supermatrix representation}

\noi The above superalgebra can be realized by the 
5 $\times$ 5 supermatrices:

  \eq
  M_{ab} = \left(
\begin{array}{cc}
 {1 \over 2} \gamma_{ab} &  0    \\
 0  &  0   \\
\end{array}
\right),  ~~ P_{a} = \left(
\begin{array}{cc}
 {1 \over 2} \gamma_{a} &  0    \\
 0  &  0   \\
\end{array}
\right)
  , ~~\Qbar_\alpha= Q^\beta C_{\beta\alpha} =  \left(
\begin{array}{cc}
0 &  \delta^\rho_\alpha    \\
 C_{\sigma\alpha} &  0   \\
\end{array} \right)\label{representation}  
  \en
  To verify the anticommutations
  (\ref{QQcomm}), one needs the identity 
   \eq
   \delta^\rho_\alpha C_{\sigma\beta} +   \delta^\rho_\beta C_{\sigma\alpha} = -{1 \over 2}   (C\gamma_{a})_{\alpha\beta} (\gamma^a)^\rho_{~\sigma} +{1 \over 4}   (C\gamma_{ab})_{\alpha\beta} (\gamma^{ab})^\rho_{~\sigma} 
       \en
 deducible from the Fierz identity (\ref{2psiFierz}) by factoring out the two spinor Majorana one-forms.
 
\subsection{Connection and curvature }

\noi The $1$-form $OSp(1|4)$-connection  is given by
\eq
\Ombold =  {1 \over 2} \omega^{ab} M_{ab}  + V^a P_a + \Qbar_\alpha \psi^\alpha
\en
In the $5 \times 5$ supermatrix representation:
\eq
  \Ombold \equiv 
\left(
\begin{array}{cc}
  \Om &  \psi    \\
 -\psibar  &  0   \\
\end{array}
\right), ~~~ \Om \equiv \unquarto \om^{ab} \ga_{ab} + {1 \over 2} V^a \ga_a
    \label{Omdef4}  
  \en
The corresponding $OSp(1|4)$ curvature two-form supermatrix is
 \eq
      \Rbold =  d \Ombold - \Ombold \we \Ombold~
  \equiv  \left(
\begin{array}{cc}
   R &  \Sigma    \\
- \Sigmabar  &  0   \\
\end{array}
\right) \label{Rdef4}
       \en
       \noindent where simple matrix algebra yields \footnote{we omit wedge products between forms.}:
   \eqa
    & & R = \unquarto R^{ab} \ga_{ab} + {1 \over 2} R^a \ga_a  \label{defR4}\\
    & & \Sigma = d \psi - \unquarto \om^{ab} \ga_{ab} \psi - {1 \over 2} V^a \ga_a \psi \label{defSigma4} \\
    & & \Sigmabar = d \psibar - \unquarto \psibar ~\om^{ab} \ga_{ab} - {1 \over 2}  \psibar V^a \ga_a \\
        & & R^{ab} \equiv d \om^{ab} - \om^{a}_{~c} ~\om^{cb} - V^a V^b - {1 \over 2} \psibar \ga^{ab} \psi \\
     & & R^{a} \equiv d V^{a} - \om^{a}_{~b} V^{b}  +  {1 \over 2} \psibar \ga^a \psi \label{defRa}
     \ena
 \noi We have also used the Fierz identity for $1$-form Majorana spinors in (\ref{2psiFierz}).

\sect{The $D=2+2$, $N=1$ (chiral) supergravity action}

\subsection{Action}

The  action is written in terms of the $OSp(1|4)$ curvature two-form $\Rbold$ as:
 \eq
     S = -2 \int STr ( \Rbold \Rbold \Gammabold)    \label{SG2+2action}
    \en
    where $STr$ is the supertrace and $\Gammabold$ is the constant matrix:
     \eq
       \Gammabold \equiv  \left(
\begin{array}{cc}
  \ga_5 &  0    \\
 0 &  1   \\
\end{array}
\right)
\label{Gammabold4}
       \en
\noi All boldface quantities are  5 $\times$ 5 supermatrices. 

 \subsection{Invariances}

 \noi Under the $OSp(1|4)$
  gauge transformations:
   \eq
   \de_{\epsibold} \Ombold = d \epsibold - \Ombold \epsibold + \epsibold \Ombold ~ \Longrightarrow~  \de_{\epsibold} \Rbold =  - \Rbold \epsibold + \epsibold \Rbold     \label{Omgaugevariation}
       \en
       where $\epsibold$ is  the $OSp(1|4)$ gauge parameter:
        \eq
         \epsibold \equiv 
         \left(
\begin{array}{cc}
 \unquarto  \epsi^{ab} \ga_{ab} + {1 \over 2} \epsi^a \ga_a &  \epsilon   \\
- \epsilonbar  &  0   \\
\end{array}
\right)   \label{gaugeparam4}
 \en
  the action (\ref{SG2+2action}) varies as
 \eq
\delta S = -2 \int STr (\Rbold \Rbold  [\Gammabold,\epsibold] ) \label{Sgaugevariation}
 \en
Computing the commutator yields
 \eq
  [\Gammabold,\epsibold] =   \left(
\begin{array}{cc}
  \epsi^{a}  \gamma_{a} \gamma_{5} &  (\gamma_{5} - 1) \epsilon   \\
 \epsilonbar (\gamma_{5} - 1)  &  0   \\
\end{array}
\right) 
\en
 \noi Thus the action is invariant under gauge variations with  $\epsi^a =0$ and  $(\gamma_{5} - 1) \epsilon=0$  (which implies also  $ \epsilonbar (\gamma_{5} - 1)=0$).  These restrictions on the gauge parameters determine a subalgebra of $OSp(1|4)$, generated by
 $M_{ab}$  and  $\Qbar_{\alpha} P_{+}$, where $P_{+} = {1 \over 2} (1+\gamma_{5})$. These generators close on the $OSp(1|2) \oplus Sp(2)$ subalgebra:
 \eqa
& & [M^\pm_{ab},M^\pm_{cd}] =  \eta_{bc} M^\pm_{ad} + \eta_{ad} M^\pm_{bc} - \eta_{ac} M^\pm_{bd} - \eta_{bd} M^\pm_{ac}  
\label{MMcommsub}\\
 & & \{ \Qbar^+_{\alpha}, \Qbar^+_{\beta} \} = {1 \over 2}
 (C\gamma^{ab})_{\alpha\beta} M^+_{ab}
  \label{QQcommsub} \\
   & & [M^+_{ab},\Qbar^+_{\beta}]={1 \over 2} ~(\gamma^+_{ab})^{\alpha}_{~\beta} \Qbar^+_\alpha \label{MQcommsub}\\
   & &  [M^-_{ab},M^+_{cd}] = [M^-_{ab},\Qbar^+_{\beta}]=0   
 \ena
 with
 \eq
 \gamma^+_{ab} = {1 + \gamma_5 \over 2} \gamma_{ab} = {1 \over 2} (\gamma_{ab} - {1 \over 2} \epsilon_{abcd} \gamma^{cd})
 \en
 and
 \eq
   M^\pm_{ab} = {1 \over 2} (M_{ab} \mp {1 \over 2} \epsilon_{abcd} M^{cd}) ,~~\Qbar^+_{\alpha} =  
 \Qbar_{\alpha} {1 + \gamma_5 \over 2}
 \en
The selfdual $M^+_{ab}$ and Weyl projected $\Qbar^+_{\alpha}$ generate $OSp(1|2)$, while the antiselfdual $M^-_{ab}$ generates $Sp(2)$.

  \subsection{$OSp(1|2) \oplus Sp(2)$ transformation laws}

\noi  Restricting the gauge parameter $\epsibold$ to the $OSp(1|2) \oplus Sp(2)$ subalgebra as described above, from (\ref{Omgaugevariation}) we deduce the transformation laws on the fields $\omega^{ab}$, $V^a$ and $\psi$ leaving the action (\ref{SG2+2action}) invariant:
 \eqa
 & & \delta \omega_+^{ab} = d \epsi_+^{ab} - \omega_+^{ac} \epsi_+^{db} \eta_{cd} +
 \omega_+^{bc} \epsi_+^{da} \eta_{cd} + \epsilonbar_+ \gamma^{ab} \psi_+\\
 & & \delta \omega_-^{ab} = d \epsi_-^{ab} - \omega_-^{ac} \epsi_-^{db} \eta_{cd} +
 \omega_-^{bc} \epsi_-^{da} \eta_{cd}  \\
 & & \delta V^a = (\epsi_+^{ab} + \epsi_-^{ab}) V^c \eta_{bc}  -  \epsilonbar_+ \gamma^a \psi_- \\
   & & \delta \psi_{+} = d \epsilon_+ - {1 \over 4 } \omega_+^{ab} \gamma_{ab} \epsilon_+ + {1 \over 4} \epsi_+^{ab} \gamma_{ab} \psi_+\\
     & & \delta \psi_{-} = - {1 \over 2 } V^a \gamma_a \epsilon_+ + {1 \over 4} \epsi_-^{ab} \gamma_{ab} \psi_-
 \ena
 where $\epsilon_+=P_+ \epsilon$ is the Weyl projected supersymmetry parameter, and $\epsi_+^{ab}$,
 $\epsi_-^{ab}$ are the selfdual and antiselfdual $SO(2,2) \approx Sp(2) \times Sp(2)$ parameters.
Moreover $\psi_+$ and $\psi_-$ are respectively Weyl and anti-Weyl projections of the Majorana gravitino, i.e. $\psi_\pm = P_\pm \psi$. 
 
 Thus we see that the $OSp(1|2) \oplus Sp(2)$ gauge fields $\omega_\pm^{ab}$, $\psi_+$ transform with the
 $OSp(1|2) \oplus Sp(2)$ covariant derivative of the gauge parameters, whereas the ``matter fields"
 $V^a$, $\psi_-$ transform homogeneously. Note also
 that gauge and matter fields do not mix, separating into a gauge and a matter multiplet under 
$OSp(1|2) \oplus Sp(2)$ transformations.

Finally, $\omega^{ab}_-$ is inert under supersymmetry, This will be important for the
consistency of the selfduality condition $\omega^{ab}_-=0$, see Section 3.6.
 
  \subsection{The action in terms of component fields}
  
  Recalling that $\int STr (\Rbold \Rbold)$ is a topological term, we have:
  \eq
   S=-2 \int STr (\Rbold \Rbold \Gammabold) = 4 \int STr (\Rbold \Rbold {1-\Gammabold \over 2})
   \en
   up to boundary terms. Carrying out the supertrace leads to:
\eq
 S = 4 \int Tr( R R P_-) + \Sigmabar P_-\Sigma 
   \en
   with $R$ and $\Sigma$ as defined in Section 2.3, and $P_- = (1 - \gamma_5 )/2$.
  After inserting the curvature definitions the action takes the form
   \eqa
     S = 
    \int \Rcal^{ab} V^c V^d \epsi_{abcd} - 4 \rhobar_+ \ga_a  \psi_- V^a -{1 \over 2} (V^a V^b V^c V^d + \psibar \ga^{ab} \psi V^c V^d ) \epsilon_{abcd} + \nonumber\\
 +{1 \over 2} R^{ab}_- ~\psibar_- \gamma^{cd} \psi_- \epsilon_{abcd}  ~~~~~\label{action2}
    \ena
  with
   \eq
    \Rcal^{ab} \equiv d \om^{ab} - \om^{a}_{~c} ~\om^{cb} , ~~\rho \equiv d \psi - \unquarto \om^{ab} \ga_{ab} \psi,~~ \rho_+ \equiv P_+ \rho
     \label{defRcalrho}
     \en
   We have dropped the topological term $\Rcal^{ab}_- \Rcal^{cd}_- \epsilon_{abcd}$ (sum of Euler and Pontryagin forms), and used the identities
  \eqa
   & & R^a R^a = - {\cal R}^{ab} V^a V^b - \rhobar \gamma_a \psi V^a  + {1 \over 2} \psibar \gamma_a \psi R^a + total~derivative\\
   & & \rho_- \rho_-= -{1 \over 4} \psibar_- \gamma_{ab} \psi_- {\cal R}_-^{ab} + total~derivative \\
  & & \psibar \gamma^a \psi R^a = -2 \rhobar \gamma^a \psi V^a + total~derivative \\
  & & R^{ab}_\pm = \mp {1 \over 2} \epsilon^{ab} _{~~cd}R_\pm^{cd}
 \ena
  and the Bianchi identities
   \eqa
    & & dR^a - \omega^a_{~b} R^b = - {\cal R}^{a}_{~b} V^b + \rhobar \gamma^a \psi \\
    & & d\rho - {1 \over 4 } \omega^{ab} \gamma_{ab} \rho = - {1 \over 4} {\cal R}^{ab} \gamma_{ab} \psi  
    \ena 
   consequences of the definitions (\ref{defRa}) and (\ref{defRcalrho}) of $R^a$ and $\rho$. 
   
   This action is similar to a MacDowell-Mansouri action in $D=2+2$, or also to $R^2$-type actions previously
   considered in the literature concerning self-dual supergravity,  but with the important
   difference that it is invariant under a {\it gauge (chiral) supersymmetry}, closing
   off-shell. 
  
\subsection{Equations of motion}

 \noi The variational equations for the action (\ref{action2}) read:
 \eqa
 & & R^{ab} V^c \epsi_{abcd} - 2 \Sigmabar_+ \gamma_d \psi_-=0 ~~~~~~~~~~~(Einstein ~eq.s)\\
 & & \gamma_a \Sigma_- V^a - \gamma_a \psi_- R^a = 0 ~~~~~~~~~~~~~~~~
 (gravitino~\psi_+)\\
 & & 4  \gamma_a \Sigma_+ V^a + R^{ab}_-  ~  \gamma^{cd} \psi_- \epsi_{abcd} = 0 ~~~~~~
 (gravitino~\psi_-)\\
 & & R^c V^d \epsi_{abcd} + {1 \over 2}\Sigmabar_- \gamma^{cd} \psi_- \epsi_{abcd} =0~~
 ~~~(torsion~ eq.) \label{torsioneq}
 \ena
These equations admit the vacuum solution $OSp(1|4)$ curvatures = 0.

\subsection{Self-dual D=2+2 supergravity}

We can impose the selfduality condition on the spin connection:
\eq
\omega_-^{ab} \equiv {1 \over 2} (\omega^{ab} + {1 \over 2} \epsi^{ab}_{~~cd} \omega^{cd}) =0
\label{sdcondition}
\en
Recalling that
\eq
{\cal R}^{ab} = {\cal R}^{ab}_+ + {\cal R}^{ab}_-
\en
with
\eq
 {\cal R}^{ab}_\pm = d \omega^{ab}_\pm - \omega^{ac}_\pm ~ \omega^{db}_\pm ~ \eta_{cd}
 \en
we can implement the selfduality condition (\ref{sdcondition}) in the action by discarding
the $ {\cal R}^{ab}_-$ component of $ {\cal R}^{ab}$ in the first term of (\ref{action2}), and the
$ {\cal R}^{ab}_-$ part of $R^{ab}_-$ in the last term. The resulting action 
is invariant under the transformations of $\omega_+^{ab},V^a,\psi_+,\psi_-$ given in
Section 3.3, with $\epsi^{ab}_-=0$, i.e. under $OSp(1|2)$ transformations whose gauge 
fields are $\omega_+^{ab}$ and $\psi_+$. Indeed the condition $\omega^{ab}_-=0$ breaks
$OSp(1|2) \times Sp(2)$ to its first factor $OSp(1|2)$. 

Second order formalism is retrieved by solving the torsion equation of motion (\ref{torsioneq}), which
for $\omega^{ab}_-=0$ allows to express $\omega^{ab}_+$ as a function of $V^a$, $\psi_+$ and $\psi_-$.

\sect{Conclusions}

We have presented a $D=2+2$ supergravity action, made out of the fields contained in the $OSp(1|4)$ connection. It is invariant
only under a subalgebra $OSp(1|2) \oplus Sp(2)$ of  $OSp(1|4)$. This closely resembles what happens for the Mac Dowell-Mansouri action in $D=3+1$: there too the supergavity fields are organized in 
a $OSp(1|4)$ connection, but the action itself is invariant only under the Lorentz subalgebra, whereas
in the present paper also  (1,0) supersymmetry survives, being part of the invariance subalgebra of the action.

A selfdual condition can be imposed on the spin connection, and breaks the
$OSp(1|2) \oplus Sp(2)$ invariance to the first factor $OSp(1|2)$.

\sk\sk
\noi {\bf Acknowledgements}
\sk
It is a pleasure to acknowledge useful discussions with Laura Andrianopoli, Paolo Aschieri, Bianca Cerchiai, Riccardo D' Auria, Mario Trigiante and Jorge Zanelli.

\appendix

\sect{D=2+2 $\Gamma$ matrices}

\noi {\bf Clifford algebra}
\eq
\{ \gamma_a,\gamma_b \}= 2 \eta_{ab}, ~~~~\eta_{ab} = (+1,+1, -1, -1),~~~~\gamma_{5}=\gamma_1 \gamma_{2} \gamma_3 \gamma_{4}, ~~~~\gamma_{5}^2 = 1
\en
\noi The $D=2+2$ charge conjugation matrix $C$ satisfies
\eq
\gamma_a^T= - C \gamma_a C^{-1} ,~~~C^T=-C,~~~[C,\gamma_5]=0
\en
so that $C\gamma_{a}$ and  $C\gamma_{ab}$ are symmetric.
\sk
\noi {\bf Duality relation}:
\eq
\gamma^{ab}= - {1 \over 2} \epsi^{abcd} \gamma_{cd} \gamma_5 
\en
\noi {\bf Fierz identity }
\sk
\noi The following Fierz identity holds for Majorana spinor one-forms ($\psibar = \psi C$):
  \eq
  \psi \psibar = \unquarto ( \psibar \ga^a \psi \ga_a - \unmezzo \psibar \ga^{ab} \psi \ga_{ab} )
  \label{2psiFierz}
  \en
\noi (to prove it, just multiply both sides by $\ga_c$ or $\ga_{cd}$ and take the  trace on spinor indices).

\end{document}